# Main-chain poly(ionic liquid)-derived nitrogen-doped micro/mesoporous carbons for $CO_2$ capture and selective aerobic oxidation of alcohols

Jiang Gong [a], Huijuan Lin [a], Konrad Grygiel [a], Jiayin Yuan [a,b,*]

[a] Department of Colloid Chemistry, Max Planck Institute of Colloids and Interfaces, Research Campus Golm, 14476, Potsdam, Germany
[b] Department of Chemistry and Biomolecular Science, and Center for Advanced Materials Processing, Clarkson University, 8 Clarkson Avenue, 13699 Potsdam, USA

*Corresponding author. E-mail address: jyuan@clarkson.edu

**Abstract**: Sustainable development and the recent fast-growing global demands for energy and functional chemicals urgently call for effective methods for $CO_2$ remediation and efficient metal-free catalysts for selective oxidation of aromatic alcohol. Herein, a unique main-chain poly(ionic liquid) (PIL) is employed as the precursor to prepare nitrogen-doped micro/mesoporous carbons *via* simultaneous carbonization and activation, which bear high yield, large specific surface area above 1700 $m^2$ $g^{-1}$ and rich nitrogen dopant. The porous carbon products deliver a high $CO_2$ adsorption capacity up to 6.2 mmol $g^{-1}$ at 273 K and 1 bar with outstanding reversibility and satisfactory selectivity. Besides, they work excellently as metal-free carbocatalysts for the selective aerobic oxidation of benzyl alcohol to benzaldehyde with high selectivity. It is believed that this work not only provides a facile approach to prepare nitrogen-doped porous carbon, but also advances the related research in the fields of environment and catalysis.

# 1. Introduction

CO$_2$ emission is an important issue in the establishment of a sustainable modern society and is currently one of the driving forces to implement the green chemistry concept. The accumulation of CO$_2$ in atmosphere is widely considered as a primary factor in global climate change. Carbon capture and sequestration technologies have been proposed to be one of solutions [1,2]. One of the industrial techniques for CO$_2$ capture is chemical sorption by aqueous solution of organic amines [3]. Although these systems can achieve a high sorption capacity, they suffer from one or more of the following drawbacks, such as solvent loss, corrosion and high energy consumption for regeneration. There is also huge interest in chemical conversion of CO$_2$ into industrial raw materials [4]. Recently, significant research efforts have been devoted to exploring porous materials with large specific surface area, *e.g.*, porous frameworks (including metal organic framework (MOF) and zeolitic imidazolate framework (ZIF)) [5-8], zeolites [9], and carbons [10-12]. Nevertheless, developing porous adsorbents that efficiently capture CO$_2$ and mitigate the dilemma remains a challenging but imperative task. In parallel, the selective aerobic oxidation of aromatic alkanes/alcohols to their corresponding aldehydes/ketones is one of the most important transformations of functional groups in organic synthesis, both for fundamental research and industrial manufacturing [13-15]. Plenty of metal-catalyzed aerobic processes, especially based on noble metals, have been studied [16,17]. Considering the scarcity and high cost of noble metals and the sustainability and rational use of resources, it would be highly desirable to explore metal-free catalysts for these reactions.

Nitrogen-doped porous carbons bearing combined micro/mesopores represent an emerging class of porous materials of versatile functionalities and tunable porosities [18-20]. The micropores in a high density packing deliver a high surface area, therefore abundant active sites for both sorption and catalytic reactions, whereas the mesopores, beside their moderate surface area, enhance mass transport and diffusion of sorbates and reagents [21-23]. Additionally, from the viewpoint of heteroatom doping, the incorporation of nitrogen into graphitic carbon networks improves the oxidation stability and basicity through conjugation between the lone electron pair of nitrogen and the $\pi$ system of carbon lattice [24]. Owing to these advantages, nitrogen-doped micro/mesoporous carbons have exhibited potential applications in many fields such as energy [25-27], catalysis [28-30], and environment [31, 32]. In particular, they are promising $CO_2$ adsorbents thanks to the basic nitrogen sites that improve affinity towards acidic $CO_2$ molecules [33]. Furthermore, previous study has found that the nitrogen sites of carbocatalysts are pivotal for the C-H bond activation, because nitrogen dopant alters local electronic structure of the adjacent carbon atoms and promotes catalytic reactivity [13]. There have already been some pioneer works in this field [34,35], nevertheless fabricating functional nitrogen-doped porous carbons in replacement of precious metal-based catalysts is actively pursued for green chemical processes.

To synthesize nitrogen-doped porous carbons, a major concern is the choice of precursor, of which the chemical structure affects the production yield, nitrogen content, graphitic structure, *etc*. The yields of common organic carbon precursors are generally below 40 wt % at high temperatures (*e.g.*, 900 °C, see Table S1), as most of organic compounds completely evaporate or decompose during high temperature carbonization. Since the first report on the conversion of poly(ionic liquid)s (PILs) into porous carbon in 2010 [36], PILs have been considered to be an important class of polymer-based carbon precursors. Compared to other polymers [37,38], structurally well-defined PILs are thermally more stable at temperatures up to 400 °C, and they contain rich heteroatoms, nitrogen in most cases, yielding heteroatom-doped carbons in good yield [39,40]. The favorable thermal stability of PILs is related to their ionic nature, aromatic nature as well as some network-forming groups [41], such as cyano group that undergoes trimerization reaction to form polytriazine network under charring conditions [42].

Herein, we report how to synthesize functional porous carbons with an unusually high yield at 900°C (47 wt % to 67 wt % dependent on the activation agent amount) from a main-chain PIL simultaneously bearing nitrogen-rich imidazolium cation, cyano group and aromatic ring *via* one-step carbonization/activation process. The as-formed nitrogen-doped carbons bearing abundant micro/mesopores not only store $CO_2$ as high as 6.2 mmol g$^{-1}$ at 273 K and 1 atm but also serve as high-performance metal-free carbocatalysts for selective aerobic oxidations, here exemplified by the conversion of benzyl alcohol to benzaldehyde.

## 2. Materials and methods

### 2.1. Materials

Glacial acetic acid (purity ≥ 99.7%, Alfa Aesar), *p*-phenylenediamine (purity ≥ 99%, Sigma-Aldrich), pyruvaldehyde (40% aqueous solution, Sigma-Aldrich), formaldehyde (37% aqueous solution, Sigma-Aldrich), sodium dicyanamide (NaN(CN)$_2$, purity ≥ 96%, Sigma-Aldrich), potassium hydroxide (KOH, purity ≥ 90%, Sigma-Aldrich), and benzyl alcohol (purity ≥ 99%, Sigma-Aldrich) were of analytical grade and used as received without further purifications. All other chemicals were utilized without further purifications.

### 2.2. Synthesis of the PIL precursor (PILPhDCA)

Firstly, an imidazolium-type PILPhAc was synthesized in water *via* one-pot modified Debus-Radziszewski reaction (Scheme 1), as we reported previously [43]. In a typical run, Milli-Q® water (400 mL) and glacial acetic acid (33.0 mL) were added under vigorous stirring to 10.0 g of *p*-phenylenediamine. Such mixture was then injected to the mixture of pyruvaldehyde (14.6 mL) and formaldehyde (7.1 mL). The solution was stirred for 15 min, diluted with Milli-Q® water and dialyzed against Milli-Q® water (the molecular weight cut-off of dialysis bag is 3.5 kDa) for 1 week. PILPhDCA was prepared by anion exchange from PILPhAc with excessive NaN(CN)$_2$ in aqueous solution. In a typical run, 4.0 g of PILPhAc was added to 800 mL of Milli-Q® water. After complete dissolution, 100 mL of an aqueous solution containing 5 molar equivalents of NaN(CN)$_2$ was added under vigorous stirring for 30 min. The precipitation occurred. The precipitate was filtered off, washed with Milli-Q® water several times, and dried at 80 °C to constant weight under high vacuum.

### 2.3. Preparation of nitrogen-doped porous carbons (NPCs)

Dried PILPhDCA powder was firstly mixed with KOH in Milli-Q® water at a KOH/PILPhDCA mass ratio of 2, 4 and 6, respectively. The mixture was agitated for 30 min to achieve a uniform suspension, and then dried at 80 °C to obtain a solid mixture of KOH and PILPhDCA. Afterwards, the solid mixture was heated at 450 °C for 1 h under nitrogen atmosphere and subsequently calcined at 900 °C for 1 h. The ramping rate was maintained at 10 °C min$^{-1}$ using a Nabertherm N7/H chamber oven with a P300 controller. After slowly cooling down to room temperature, the product was repeatedly washed with 1 mol L$^{-1}$ HCl and Milli-Q® water until the pH value of the filtrate reached 7 ± 0.5 to ensure the efficient removal of KOH (< 0.05 wt% in the final carbon products), and finally dried at 110 °C under vacuum for 12 h. The resultant activated carbon sample was named as NPC-2, NPC-4 and NPC-6, respectively, while the carbon product prepared from merely PILPhDCA in the absence of KOH under the similar preparation process was denoted as NPC-0.

## 2.4. Characterization

Proton nuclear magnetic resonance ($^1$H NMR) spectra were recorded at room temperature on a VARIAN 400-MR (400 MHz) spectrometer using DMSO-$d_6$ as the solvent. Attenuated total reflection (ATR) Fourier-transform infrared spectroscopy (FTIR) was performed at room temperature on a BioRad 6000 FT-IR spectrometer equipped with a Single Reflection Diamond ATR. Gel permeation chromatography (GPC) was performed by using NOVEMA-column with a mixture of 80% of acetate buffer and 20% of methanol (flow rate = 1.0 mL min$^{-1}$, dextran standards using RI detector-RI-101 Refractometer). Thermogravimetric analysis (TGA) experiment was performed under nitrogen flow at a heating rate of 10 °C min$^{-1}$ using a Netzsch TG209-F1 apparatus. Nitrogen adsorption/desorption experiments were performed with a Quantachrome Autosorb and Quadrasorb at 77 K, and the data were analyzed using Quantachrome software. The specific surface area was calculated using the Brunauer-Emmett-Teller (BET) equation. The samples were degassed at 150 °C for 24 h before measurements. Combustion elemental analyses were done with a varioMicro elemental analysis instrument from Elementar Analysensysteme. The surface element compositions were characterized by means of X-ray photoelectron spectroscopy (XPS) carried out on a VG ESCALAB MK II spectrometer using an Al K$α$ exciting radiation from an X-ray source operated at 10.0 kV and 10 mA. Scanning electron microscopy (SEM) measurements were carried out in a LEO 1550-Gemini electron microscope (acceleration voltage = 3 kV), and the samples were coated with a thin gold layer before SEM measurements. Energy-dispersive X-ray (EDX) map was taken on the SEM with an EDX spectrometer. Transmission electron microcopy (TEM) measurements were performed using a Zeiss EM 912 (acceleration voltage = 120 kV). High-resolution TEM (HRTEM) measurement was carried out on a FEI Tecnai G2 S-Twin transmission electron microscope operating at 200 kV. X-ray diffraction (XRD) pattern was recorded on a Bruker D8 diffractometer using Cu K$\alpha$ radiation ($\lambda$ = 0.154 nm) and a scintillation counter. Raman spectrum was collected using a confocal Raman microscope ($\alpha$300; WITec, Ulm, Germany) equipped with a 532 nm laser. Gas chromatography-mass spectrometry (GC-MS) analyses were performed using an Agilent Technologies 5975 gas chromatograph equipped with a MS detector and a capillary column (HP-5MS, 30 m × 0.25 mm × 0.25 μm). The temperature program started with an isothermal step at 50 °C for 2 min, in a second step the temperature was increased to 300 °C at a rate of 30 °C min$^{-1}$ and then kept for 1 min.

## 2.5. $CO_2$ capture

$CO_2$ capture by NPCs was conducted by measuring the adsorption isotherms at 0 and 25 °C in a Quantachrome Autosorb and Quadrasorb (Quantachrome Instruments). NPCs were firstly degassed at 150 °C for 30 h before each analysis. The isosteric heat of adsorption ($Q_{st}$) was calculated by applying the Clausius-Clapeyron equation to the adsorption isotherms measured at the aforementioned two temperatures:

$$\ln(\frac{p_1}{p_2}) = Q_{st} \times \frac{T_2 - T_1}{R \times T_1 \times T_2} \qquad (1)$$

where $R$ is the universal gas constant (8.314 J mol$^{-1}$ K$^{-1}$), $p_1$ and $p_2$ denote the pressure (Pa), and $T_1$ and $T_2$ represent the absolute temperature (K).

The breakthrough experiment of NPC-4 was performed in a small-scale fixed-bed column. The feed stream with a composition of $CO_2/N_2$ (20% of $CO_2$ in volume) was fed into the column at a flow rate of 20 mL min$^{-1}$. Prior to the sorption experiment, the sample was heated to 200 °C under $N_2$ flow at 100 mL min$^{-1}$ for 2 h to desorb adventitious $CO_2$ and water, slowly cooled to 25 °C and then exposed to $CO_2$ for the experimental sorption run.

## 2.6. Selective aerobic oxidation of benzyl alcohol

The aerobic oxidation reaction was carried out in Schlenk tube containing 40 mg of catalyst (optimized amount), 100 mg of benzyl alcohol, and a stir bar at 60 (or 100) °C under a static $O_2$ atmosphere (1 atm). After the reaction was completed, the reaction mixture was centrifuged at 8000 rpm for 2 min to separate the catalyst for the later analyses of GC-MS and $^1$H NMR.

## 3. Results and discussion

### 3.1. Synthesis and characterization of the PIL precursor

The synthetic route to the main-chain PILPhDCA is depicted in Scheme 1, based on our recent work [43]. An imidazolium-type PILPhAc was firstly synthesized *via* one-pot modified Debus-Radziszewski reaction in water at room temperature. Briefly, water and glacial acetic acid were added to *p*-phenylenediamine under vigorous stirring. The solution was injected to a mixture of pyruvaldehyde and formaldehyde to prepare PILPhAc. PILPhDCA was obtained by anion exchange of PILPhAc with excessive $NaN(CN)_2$ in aqueous solution. The successful synthesis of PILPhDCA is verified by $^1$H NMR, FTIR, and GPC measurements (Figs. S1−S3).

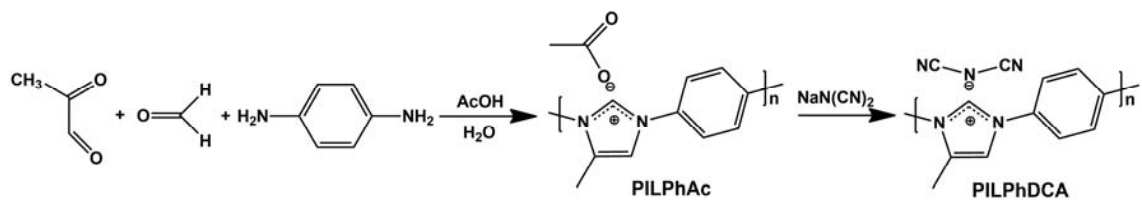

**Scheme 1** Schematic illustration of the PILPhDCA synthesis *via* one-pot modified Debus-Radziszewski reaction in water at room temperature.

The thermal stability of PILPhDCA was studied by TGA under nitrogen atmosphere (Figs. 1 and S4). The temperature at the maximum weight loss rate ranges from 360 to 510 °C, which is mainly attributed to the fractionation during the crosslinking reaction of the cyano group [43]. Surprisingly, the carbonization yields at 600 and 900 °C according to the TGA analysis are 73 and 66 wt %, respectively. The unusually high carbonization yield of PILPhDCA is attributed to its special chemical structure bearing both cyano group to form stable networks and aromatic backbone systems that end up in the final carbon product with less weight loss.

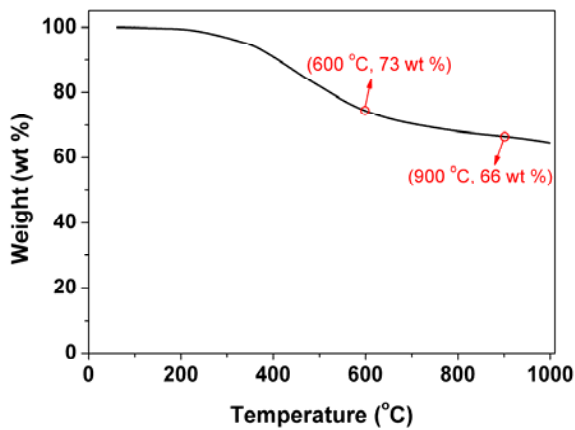

**Fig. 1.** TGA curve of PILPhDCA under nitrogen atmosphere at 10 °C min$^{-1}$.

It is well known that ionic liquid and PIL are not volatile, but at elevated temperatures they decompose partially into volatile products. As a result, during heating up to 900 °C under inert gas atmosphere, most PILs decompose nearly completely or leave a small amount of residual mass up to 40 wt % [31,42]. The results above highlight the excellent carbonization function of PILPhDCA as a new precursor for the preparation of nitrogen-doped carbon, which is superior to many common nitrogen-containing polymers (Table S1), *e.g.*, polyaniline [44] and polyacrylonitrile [45].

### 3.2. Yield, textural property, and element composition of NPCs

Considering the fascinating carbonization yield of PILPhDCA in a bulk state, we modified the carbonization process by adding KOH as activation agent to PILPhDCA at a KOH/PILPhDCA mass ratio of 2, 4 and 6 to prepare nitrogen-doped porous carbons (NPCs). The as-synthesized carbon products after thorough water rinsing are denoted as NPC-2, NPC-4 and NPC-6, respectively, while NPC-0 denotes carbon prepared without KOH. Table 1 presents the yield of NPCs obtained at different KOH/PILPhDCA mass ratios. NPC-0 shows an ultrahigh yield of 67 wt % at 900 °C, which is in good agreement with the TGA result (66 wt % in Fig. 1). When the KOH/PILPhDCA mass ratio increases to 2, the resultant

NPC-2 still exhibits a high yield of 58 wt %. The decrease in carbonization yield is attributed to the etching of the as-formed graphitic layers by KOH at the carbonization temperature to generate porous structure, which will be clarified in detail in the following part. Further increasing KOH/PILPhDCA mass ratio leads to a gradual decrease in the carbonization yield of NPCs, *e.g.*, 54 wt % for NPC-4 and 47 wt % for NPC-6. Nevertheless they are significantly high in comparison to many polymeric precursors. The superior carbonization yield and easy implementation enable a large-scale production of NPCs, for example in a 40 g scale (Fig. S5), the largest capacity allowed by our laboratory carbonization oven.

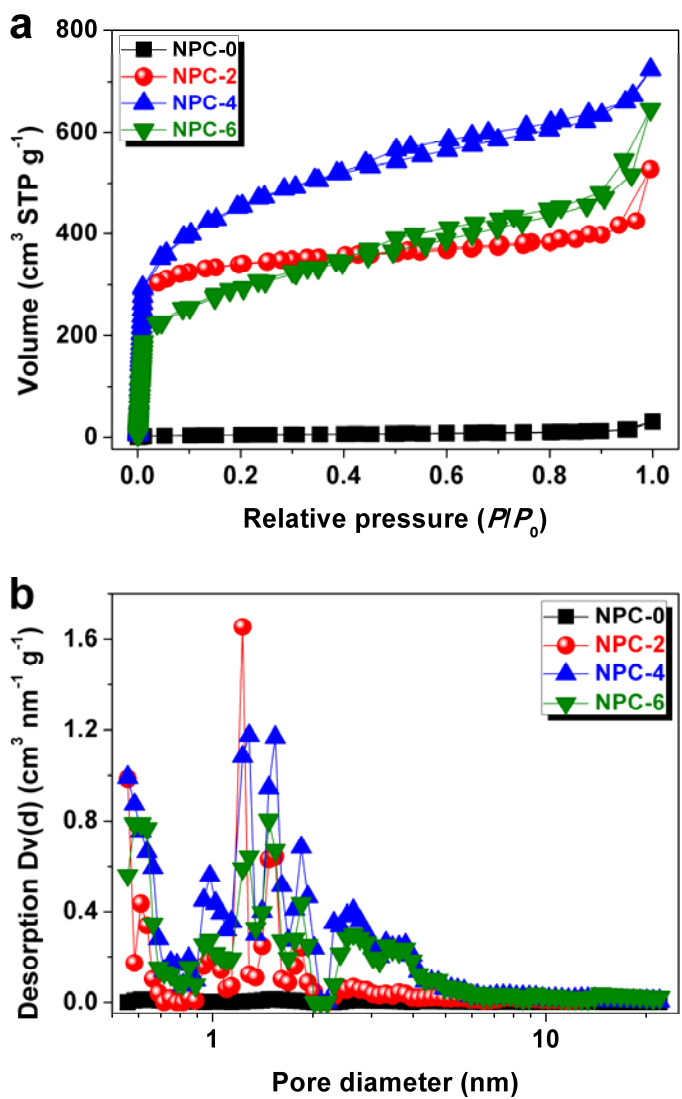

**Fig. 2.** (a) Nitrogen adsorption/desorption isotherms at 77 K and (b) pore size distribution plots of NPCs prepared at different KOH/PILPhDCA mass ratios.

The porosity of carbon materials was probed using nitrogen gas sorption at 77 K. Fig. 2 shows the nitrogen adsorption/desorption isotherms and pore size distribution plots of the as-prepared NPCs. NPC-0 displays a negligible specific surface area ($S_{total}$) of 17 m$^2$ g$^{-1}$ and low pore volume ($V_{total}$) of 0.024 cm$^3$ g$^{-1}$ (Table 1), that is to say, a poorly porous carbon. In comparison, the nitrogen adsorption/desorption isotherms of NPC-2 exhibit type I isotherm with a significant adsorption at the relative pressure $P/P_0 < 0.1$ due to the capillary filling of micropores, revealing the presence of rich micropores. The $S_{total}$ and $V_{total}$ of NPC-2 are determined to be 1216 m$^2$ g$^{-1}$ and 0.963 cm$^3$ g$^{-1}$, respectively. Compared with NPC-0, the substantial $S_{total}$ and $V_{total}$ enhancements of NPC-2 are obviously attributed to KOH activation. The activation mechanism is normally suggested to be the following reaction [46]:

$$6KOH + 2C \rightarrow 2K + 3H_2 + 2K_2CO_3 \tag{2}$$

When the temperature is higher than 700 °C, the reaction proceeds as follows:

$$K_2CO_3 + C \rightarrow K_2O + 2CO \tag{3}$$

$$K_2CO_3 \rightarrow K_2O + CO_2 \tag{4}$$

$$2K + CO_2 \rightarrow K_2O + CO \tag{5}$$

When the temperature is higher than 800 °C, it can be expressed as:

$$K_2O + C \rightarrow 2K + CO \tag{6}$$

With the increase of KOH/PILPhDCA mass ratio, the obtained NPC-4 and NPC-6 display type I/IV isotherms. A higher adsorption capacity is observed at the low relative pressure $P/P_0 < 0.1$, meaning the presence of more micropores. Besides, a detectable type-H4 hysteresis loop at the relative pressure $P/P_0$ ranging from 0.4 to 0.8 is found, corresponding to the filling and emptying of mesopores by capillary condensation. This result implies that the addition of more KOH is propitious to the formation of both micropores and mesopores. NPC-4 thus exhibits the highest $S_{total}$ (1742 m$^2$ g$^{-1}$) and largest $V_{total}$ (1.415 cm$^3$ g$^{-1}$). After further increasing the amount of KOH, the micropores are prone to merge into mesopores and/or macropores; NPC-6 thereupon displays relatively lower $S_{total}$ (1141 m$^2$ g$^{-1}$) and $V_{total}$ (1.063 cm$^3$ g$^{-1}$) than NPC-4, but still higher than NPC-0.

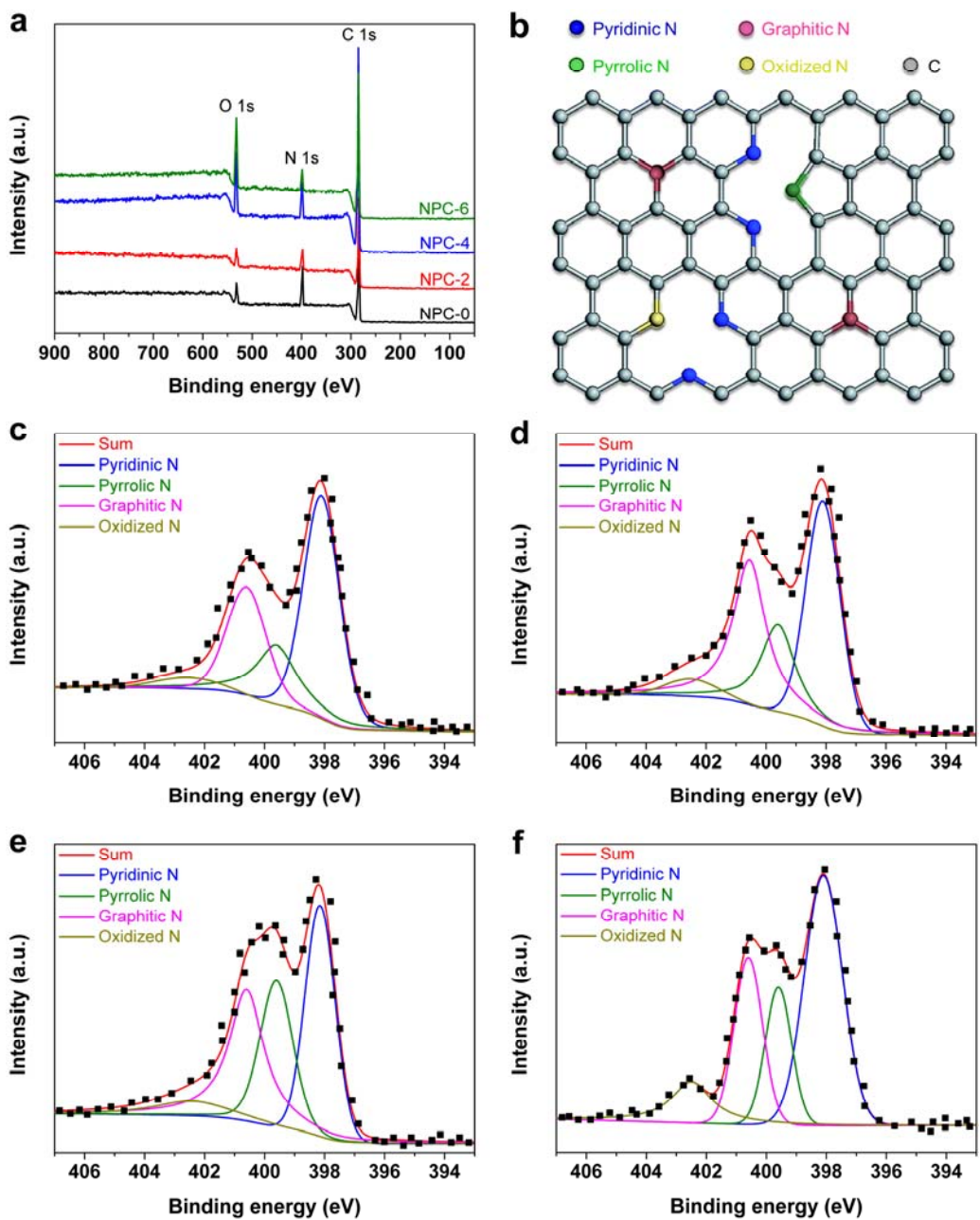

**Fig. 3.** (a) XPS spectra of NPCs, (b) scheme of nitrogen element with different states, and high-resolution N 1s XPS spectra of (c) NPC-0, (d) NPC-2, (e) NPC-4, and (f) NPC-6.

Combustion element analyses were carried out to access the element compositions of NPCs (Table S2). Generally, they consist of carbon (79.6–82.0 wt %), nitrogen (3.7–10.7 wt %) and oxygen (7.0–14.9 wt %) with a trace amount of hydrogen (1.6–2.0 wt %). Upon the increase of KOH/PILPhDCA mass ratio, the nitrogen content decreases while oxygen content increases, which are consistent with the previous study [47]. The surface element compositions of NPCs were investigated by using XPS. The full XPS survey spectra of NPCs show the expected presences of carbon, nitrogen and oxygen (Fig. 3a). Their compositions were calculated from the corresponding peak areas of XPS spectra (Table S3). Curve deconvolution shows that the high-resolution N 1s spectra of the samples can be well-fitted to the superposition of four peaks. The peaks with binding energies centered at 398.4, 400.4, 401.3 and 404.6 eV are assigned to pyridinic, pyrrolic, graphitic and oxidized N, respectively (Fig. 3b). Pyridinic N and graphitic N are the dominant nitrogen-containing surface functional groups in all samples, with the contents in the ranges of 33.7–48.2% and 23.8–35.8%, respectively (Figs. 3c–3f, Table S3). Pyrrolic N and oxidized N cannot be neglected, with the abundance in the range of 17.8–25.4% and 4.5–12.2%, respectively.

### 3.3. Morphology, microstructure, and phase structure of NPCs

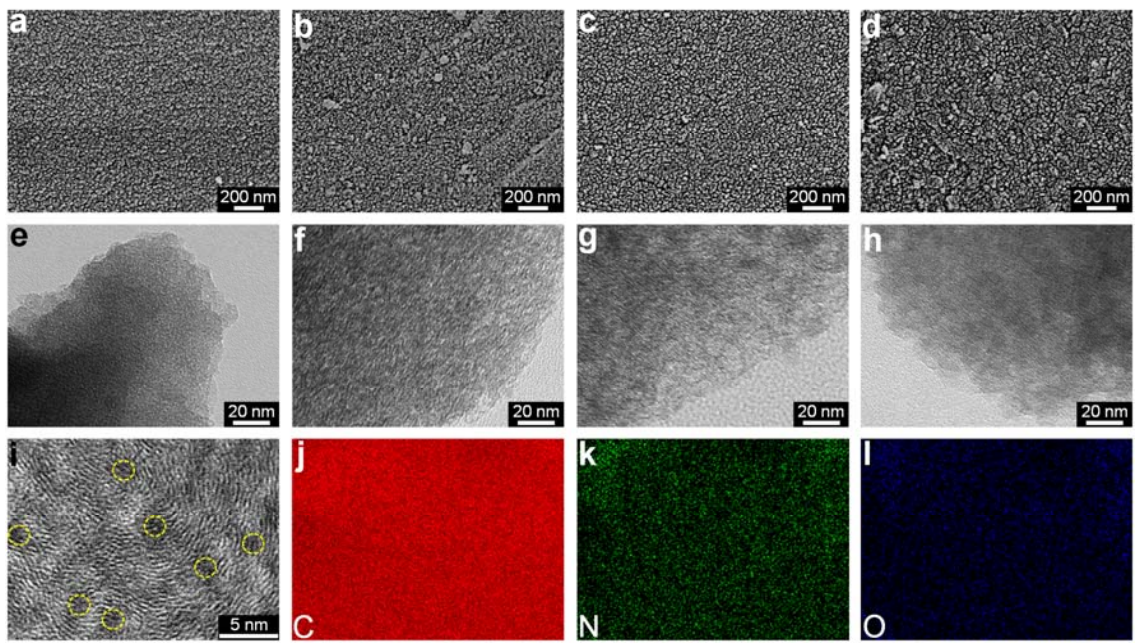

**Fig. 4.** SEM images of (a) NPC-0, (b) NPC-2, (c) NPC-4 and (d) NPC-6 with a high magnification. TEM images of (e) NPC-0, (f) NPC-2, (g) NPC-4 and (h) NPC-6. (i) HRTEM image of NPC-4; the yellow cycles show the nanoscale pores. EDX maps of (j) carbon, (k) nitrogen, and (l) oxygen for NPC-4 according to SEM image (c).
13

Fig. 4 displays SEM and TEM images of typical areas of the NPC-*x* samples. With the mass ratio of KOH to PILPhDCA increasing from 0 to 6, the bulk morphologies of NPCs display no obvious differences (Fig. S6), being composed of dense patch-like particles with a size ranging from 10 to 40 nm (Figs. 4a–4d). To observe the local microstructure of NPCs, TEM characterization was carried out. A large number of nanopores are found in NPCs except NPC-0 (Figs. 4e–4h). HRTEM was conducted to observe the subnanometer structure of the NPC-4 product (Fig. 4i). The dark patterns are stacked bent graphitic layers in a nanoscale size, indicating a short-range order of graphitic crystallite. Pores of few nanometers (pointed by cycles) were also observed in the NPC-4, similar to the porous graphene activated by KOH as reported previously [48]. EDX maps of carbon, nitrogen and oxygen (Figs. 4j–4l) certify that the NPC-4 sample is uniform with respect to element distributions.

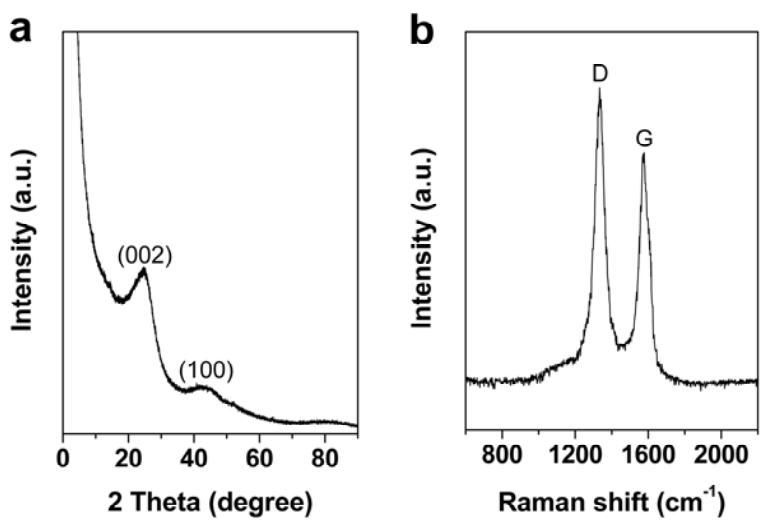

**Fig. 5.** (a) XRD pattern and (b) Raman spectrum of NPC-4.

To investigate the phase structure of NPCs, XRD characterization was employed (Fig. 5a). The appearances of two broad diffraction peaks at $2\theta = 25°$ and $42°$, which are assigned to the typical graphitic (002) and (100) planes, respectively, reveal the turbostratic structure of NPC-4. Raman spectroscopy was used to gain more insights into the phase structure information of NPC-4 (Fig. 5b). The D band at ~1333 cm$^{-1}$ and G band at ~1578 cm$^{-1}$ are related to the disordered and defective structure of carbon and the ordered carbon structure with $sp^2$ electronic configuration, respectively [49]. The relatively low $I_G/I_D$ value (~0.61) indicates the partial graphitization of NPC-4, typical for microporous carbons, which is caused by the defects generated by both nitrogen doping and KOH activation.

**3.4. $CO_2$ capture by NPCs**

The $CO_2$ adsorption isotherms of the as-prepared NPCs are shown in Fig. 6a. In spite of rather low specific surface area (17 $m^2$ $g^{-1}$), NPC-0 exhibits a moderate $CO_2$ uptake of 2.0 mmol $g^{-1}$ at 273 K and 1 bar. This is no doubt due to the incorporation of abundant nitrogen functionalities (10.7 wt %), which act as basic sites for fastening acidic $CO_2$ molecules. The $CO_2$ adsorption capacity dramatically grows up to 5.0 mmol $g^{-1}$ for NPC-2 and 6.2 mmol $g^{-1}$ for NPC-4, and then declines to 4.3 mmol $g^{-1}$ for NPC-6. The ultra-high specific surface area, large pore volume (particularly the narrow micropores *via* pore filling mechanism) and rich nitrogen dopants (especially the pyridinic N and pyrrolic N due to acid-base interaction) of NPC-2 and NPC-4 mainly contribute to the significant improvement of $CO_2$ uptake. The decreasing $CO_2$ uptake by NPC-6 in comparison to NPC-2 and NPC-4 is reasonable because of the decreased specific surface area and nitrogen dopant. Moreover, the $CO_2$ uptake by NPC-4 at 298 K and 1 bar is measured to be 4.5 mmol $g^{-1}$, which is 19.8 wt % of the porous carbon (Fig. 6b). These results are among the top values reported for porous adsorbents [33,37,50-65] (see Table 2).

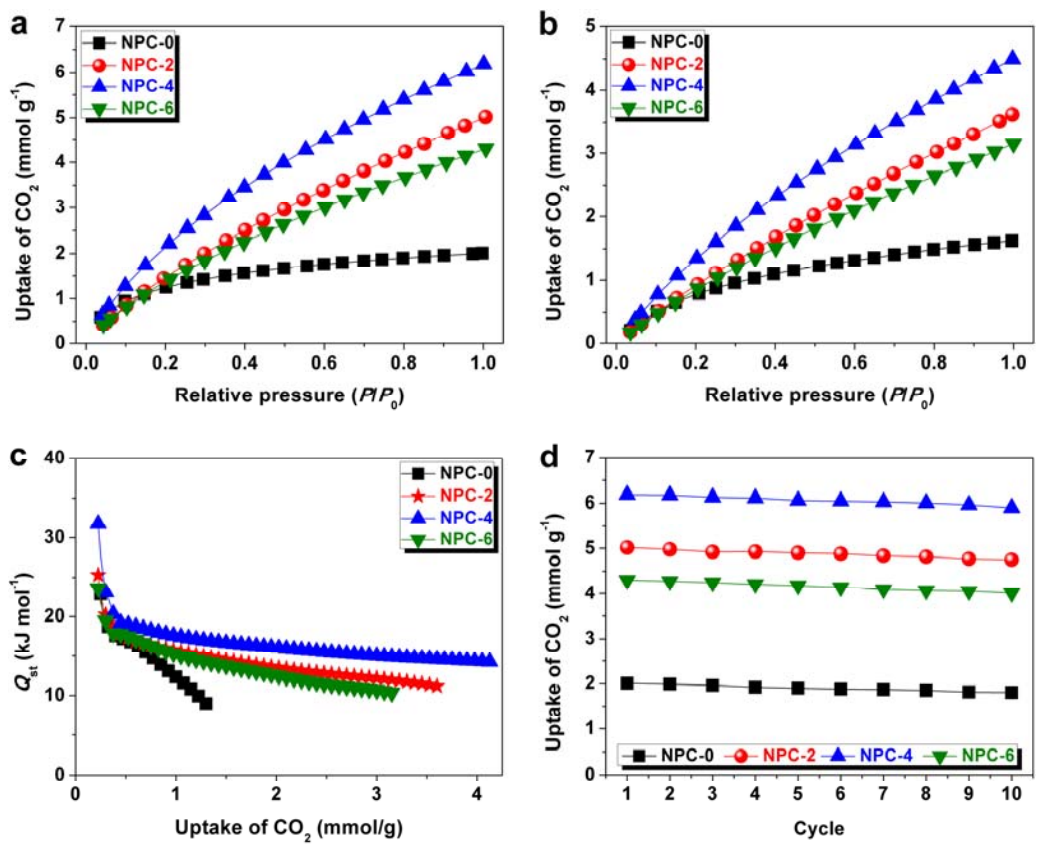

**Fig. 6.** $CO_2$ adsorption isotherms of NPCs measured at (a) 273 K and (b) 298 K. (c) Isosteric heats of $CO_2$ adsorption for NPCs at different $CO_2$ uptake amounts. (d) The reusability of NPCs for $CO_2$ uptake over 10 cycles at 273 K.

Under similar experimental conditions, the $N_2$ adsorption capacity of NPC-4 is measured to be 0.45 and 0.35 mmol g$^{-1}$ in maximum at 273 and 298 K, respectively (Fig. S7), which are considerably lower than that for $CO_2$. The initial slopes of $CO_2$ and $N_2$ adsorption isotherms are calculated, and the ratio of these slopes is applied to estimate the adsorption selectivity for $CO_2$ over $N_2$. As a result, the apparent $CO_2$/$N_2$ selectivity of NPC-4 at 273 and 298 K is determined to be 14 and 17, respectively, indicating that NPC-4 could be a potential selective adsorbent for $CO_2$/$N_2$ separation. Additionally, by fitting the $CO_2$ adsorption isotherms of NPCs at 273 and 298 K, the isosteric heats of $CO_2$ adsorption ($Q_{st}$) are calculated by using Clausius-Clapeyron equation. The calculated $Q_{st}$ values (Fig. 6c) are in the ranges of 8–23 kJ mol$^{-1}$ for NPC-0, 11–25 kJ mol$^{-1}$ for NPC-2, 15–32 kJ mol$^{-1}$ for NPC-4, and 10–24 kJ mol$^{-1}$ for NPC-6, depending on different $CO_2$ uptake amounts. Evidently,

NPC-4 shows higher $Q_{st}$ than other NPCs. These values are comparable to, if not higher than those of the previously reported adsorbents [62,63], which is probably owing to the significant presence of micropores and the strong interaction between basic nitrogen species and acidic $CO_2$ molecules. Furthermore, the reusability of NPCs for $CO_2$ capture at 273 K is tested over 10 cycles (Fig. 6d). The $CO_2$ adsorption capacity of NPC-4 after 10 cycles retains up to 5.9 mmol g$^{-1}$, higher than that of other NPCs, which indicates the good recyclability of NPC-4 for $CO_2$ capture.

To assess the potential of the sorbent in practical processes, more realistic conditions are required, *i.e.*, competitive $CO_2$ adsorption with $N_2$ in a dynamic system (Fig. S8). In a typical experiment, a mixed gas stream of 20% (v/v) $CO_2$ + 80% (v/v) $N_2$ was used to approximately simulate a postcombustion flue gas. At 298 K and 0.2 bar partial pressure of $CO_2$, the dynamic $CO_2$ capacity of NPC-4 is 1.25 mmol g$^{-1}$, which matches with that from the equilibrium measurement using pure $CO_2$ at 298 K and 0.2 bar, *i.e.*, 1.35 mmol g$^{-1}$. This implies that $CO_2$ preferentially adsorbs onto the sorbent over $N_2$, even in a $CO_2/N_2$ mixture.

### 3.5. Selective aerobic oxidation of benzyl alcohol to benzaldehyde by NPCs

It is widely acknowledged that the nitrogen dopant promotes the catalytic reactivity of carbocatalysts *via* altering the electronic structure of the adjacent carbon atoms, meanwhile the mesopores enhance the mass transport and diffusion of reagents. Consequently, nitrogen-doped micro/mesoporous carbons can be potential metal-free carbocatalysts for selective aerobic oxidations, and replace conventional base, metal and metal oxide catalysts. Here, to exemplify the application of NPCs as metal-free carbocatalysts, the selective aerobic oxidation of benzyl alcohol to higher value-added product benzaldehyde is conducted as a model reaction. Benzaldehyde has tremendous applications in perfumery, pharmaceutical, dyestuff and agrochemical industries. Commercially, benzaldehyde is synthesized by the hydrolysis of benzyl chloride or the vapor/liquid-phase oxidation of toluene. However, in the former process, traces of chlorine inevitably exist in the product benzaldehyde, while in the latter case, the selectivity to benzaldehyde is very poor.

Table 3 lists the activity results obtained with a range of reaction conditions. As the reaction temperature increases from 60 to 100 °C, the conversion of benzyl alcohol catalyzed by NPC-4 after 24 h grows up from 58.8% to 99.5%, while the selectivity keeps at 100% (entries 1 and 2). When the reaction time decreases from 24 to 12 h, a high benzyl alcohol conversion of 73.1% is retained with a selectivity of 100% (entry 3). In the control experiment using NPC-0 as catalyst, a negligible amount (0.5%) of benzyl alcohol is selectively oxidized to benzaldehyde after 24 h (entry 4), implying that apart from nitrogen dopant, the pore architecture is crucial for determining the catalytic performance. Previous report demonstrates that the sorption and activation of molecular oxygen over the graphitic nitrogen active sites to form a $sp^2$ N-$O_2$ adduct transition state is a key step [66]. Here, the high specific surface area and large pore volume of NPC-4 facilitate the easy accessibility of these active sites by molecular oxygen and benzyl alcohol, thus being beneficial for their sorption and mass transport in catalytic reaction. As an additional proof, in the control experiments using NPC-2 and NPC-6 as carbocatalysts (entries 5 and 6), the conversion of benzyl alcohol is 93.2% and 87.5%, respectively, which is lower than NPC-4 but apparently far beyond NPC-0.

To evaluate the reusability, the used NPC-4 was separated by centrifuge after the first oxidation run and employed for the next run under the same condition. After three runs (entries 7 and 8), the decreases of alcohol conversion and selectivity are negligible, suggesting the good reusability of NPC-4. It is worth pointing out that NPC-4 prevails over the previous catalysts (entries 9−14), including N-doped graphene [14], P-doped porous carbon [67], graphene oxide [68], N/O/S-doped porous carbon [69], and carbon nitride [15].

## 4. Conclusions

In summary, a novel main-chain poly(ionic liquid) bearing both cyano group and aromatic backbone conjugates have been applied as nitrogen-rich carbon precursor with unusually high carbonization yield at 900 °C (47 wt % to 67 wt %). Through simultaneous carbonization and activation of the poly(ionic liquid), nitrogen-doped micro/mesoporous carbons were prepared, which display large specific surface area up to 1742 $m^2$ $g^{-1}$ and rich nitrogen dopant. They deliver an unprecedented high $CO_2$ uptake with satisfactory selectivity and outstanding reversibility. Equally important, they serve excellently as metal-free carbocatalysts for the selective aerobic oxidation of benzyl alcohol to benzaldehyde with high conversion and selectivity. We believe that this work not only opens up a new avenue to synthesize appealing heteroatom-doped porous carbons in a high yield by rational architecture design of the polymeric precursor, but also greatly advances the related research in the fields of environment remediation and heterogeneous catalysis.

**Acknowledgements**

This work was supported by the Max Planck Society for financialsupport. J.G. thanks Prof. Tao Tang for XPS measurements and Mr.Max Braun for GC-MS measurements. J.Y. and J.G. thank the finan-cial support from the European Research Council (ERC) StartingGrant with project number 639720–NAPOLI.

*Supplementary Material (SM) for*

# Main-chain poly(ionic liquid)-derived nitrogen-doped micro/mesoporous carbons for CO$_2$ capture and selective aerobic oxidation of alcohols

Jiang Gong [a], Huijuan Lin [a], Konrad Grygiel [a], Jiayin Yuan [a,b],*

[a] Department of Colloid Chemistry, Max Planck Institute of Colloids and Interfaces, Research Campus Golm, D-14476, Potsdam, Germany
[b] Department of Chemistry and Biomolecular Science, and Center for Advanced Materials Processing, Clarkson University, 8 Clarkson Avenue, 13699 Potsdam, USA

*Corresponding author. E-mail address: jiayin.yuan@mpikg.mpg.de; jyuan@clarkson.edu

**Table S1** Comparison of the yield of nitrogen-doped carbons derived from different precursors according to the previously reported work.

| Entry | Precursor | Temperature (°C) | Yield (wt %) | Ref. in *SM* |
|---|---|---|---|---|
| 1 | Poly(ionic liquid) (PIL) | 1000 | 29.7 | [S1] |
| 2 | PIL | 1000 | 29 | [S2] |
| 3 | PIL | 1000 | 30 | [S3] |
| 4 | PIL | 1000 | 20 | [S4] |
| 5 | PIL | 800 | 24 | [S4] |
| 6 | PIL | 1000 | 21.4 | [S5] |
| 7 | PIL | 800 | 20.6 | [S5] |
| 8 | PIL | 750 | 30 | [S6] |
| 9 | Ionic liquid (IL) | 800 | 9.5–53 | [S7] |
| 10 | IL | 800 | 44.2 | [S8] |
| 11 | IL | 1000 | 21.8 | [S9] |
| 12 | IL | 1000 | 10 | [S10] |
| 13 | Protic salts | 1000 | 46 | [S11] |
| 14 | Protic salts | 100 | 36.5 | [S12] |
| 15 | Protic salts | 900 | 36.1 | [S13] |
| 16 | Protic salts | 1000 | 13 | [S14] |
| 17 | Protic salts | 800 | 23.3 | [S15] |
| 18 | Polybenzoxazine | 800 | 50–61 | [S16] |
| 19 | Polyaniline | 800 | 45 | [S17] |
| 20 | Polypyrrole | 800 | 35 | [S18] |
| 21 | Polyacrylonitrile | 1000 | 29.2 | [S19] |
| 22 | **PILPhDCA** | **600** | **73** | **This work** |
| 23 | **PILPhDCA** | **800** | **68** | **This work** |
| 24 | **PILPhDCA** | **900** | **66** | **This work** |
| 25 | **PILPhDCA** | **1000** | **64** | **This work** |

**Table S2** Element composition of NPCs.

| Entry | Sample | C [a] (wt %) | N [a] (wt %) | H [a] (wt %) | O [b] (wt %) |
|---|---|---|---|---|---|
| 1 | NPC-0 | 80.7 | 10.7 | 1.6 | 7.0 |
| 2 | NPC-2 | 82.0 | 7.2 | 1.7 | 9.1 |
| 3 | NPC-4 | 80.9 | 5.4 | 1.9 | 12.2 |
| 4 | NPC-6 | 79.6 | 3.7 | 2.0 | 14.9 |

[a] Measured by combustion element analyses. [b] Calculated by the difference.

**Table S3** Surface element composition of NPCs measured by using XPS.

| Entry | Sample | C (at %) | O (at %) | N (at %) | Pyridinic N (%) | Pyrrolic N (%) | Graphitic N (%) | Oxidized N (%) |
|---|---|---|---|---|---|---|---|---|
| 1 | NPC-0 | 79.8 | 8.1 | 12.1 | 48.2 | 20.6 | 26.7 | 4.5 |
| 2 | NPC-2 | 81.8 | 9.9 | 8.3 | 38.9 | 20.9 | 35.8 | 4.5 |
| 3 | NPC-4 | 80.1 | 13.7 | 6.2 | 33.7 | 25.4 | 33.4 | 7.5 |
| 4 | NPC-6 | 79.7 | 15.8 | 4.5 | 46.7 | 17.8 | 23.8 | 12.2 |

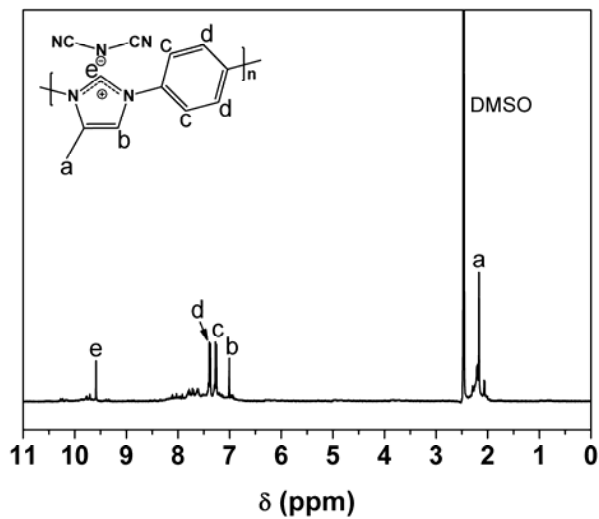

**Fig. S1.** $^1$H NMR spectrum of PILPhDCA using DMSO-$d_6$ as the solvent.

The proton signals of >C-*CH$_3$*, >C=*CH*-N<, and >N=*CH*-N< in the imidazole ring appear at 2.2, 7.0 and 9.6 ppm, respectively, while the phenyl protons show signals at 7.3 and 7.4 ppm, respectively.

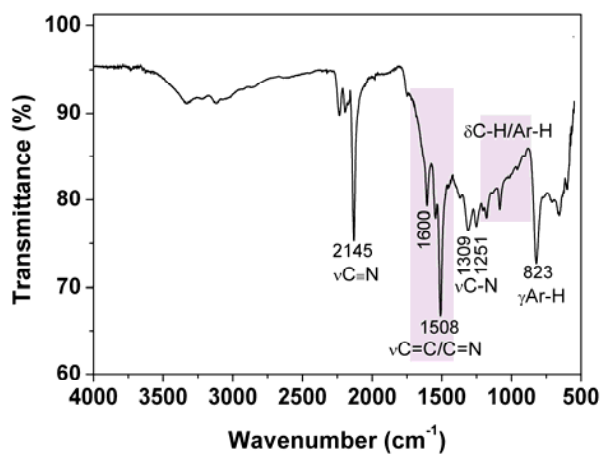

**Fig. S2.** FTIR spectrum of PILPhDCA.

The band at 2145 cm$^{-1}$ is assigned to the C≡N stretching vibration. Strong absorptions due to C=C/C=N stretching vibrations found at 1508 and 1600 cm$^{-1}$ reveal the presence of imidazole ring, which is confirmed by the imidazole ring-breathing mode at around 1251 and 1309 cm$^{-1}$. The in-plane C-H deformation of both benzene ring and imidazole ring is found in the region of 900–1200 cm$^{-1}$. Additionally, in the range of 800–900 cm$^{-1}$ we can clearly observe the out-of-plane deformation of the substituted benzene rings.

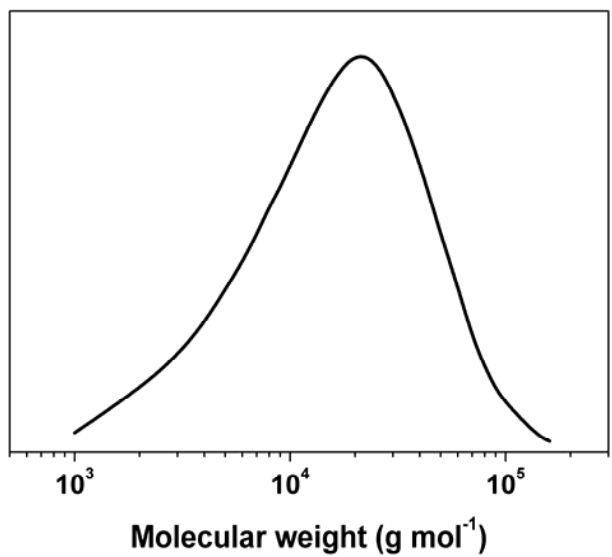

**Fig. S3.** GPC curve of PILPhAc.

Generally, GPC measurement indicates that the polymeric structures are successfully obtained during the performed reactions. The polymeric nature of the dialyzed PILPhAc was thus investigated by GPC. The synthesis of PILPhAc was halted at the early stage to access the product which retains soluble in water. The apparent number-average molecular weight of such PILPhAc can still reach $2 \times 10^4$ g mol$^{-1}$. PILPhDCA was obtained by anion exchange from PILPhAc with excessive NaN(CN)$_2$ in aqueous solution. In this case, the apparent number-average molecular weight of such PILPhDCA is determined to be ca. $2.1 \times 10^4$ g mol$^{-1}$. It should be pointed out that the apparent number-average molecular weight of PILPhDCA for the preparation of NPCs is higher than $2.1 \times 10^4$ g mol$^{-1}$, since the real synthesis of PILPhAc was not halted at early stage but to a full end of the reaction.

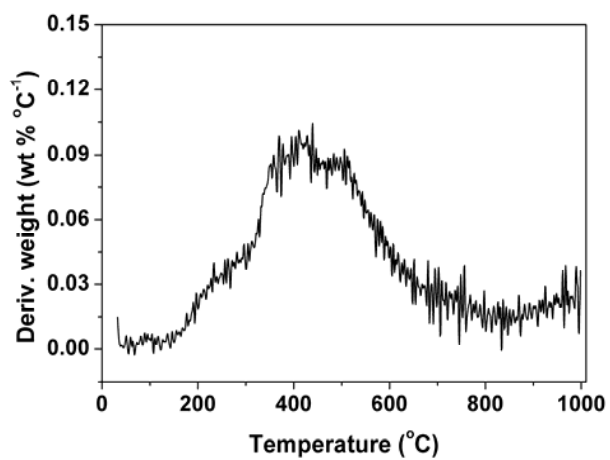

**Fig. S4.** DTG (the first derivative of TGA curve) curve for PILPhDCA under nitrogen atmosphere at 10 °C min$^{-1}$. As we can see, the temperature of the maximum weight loss rate is in the range of 360−510 °C.

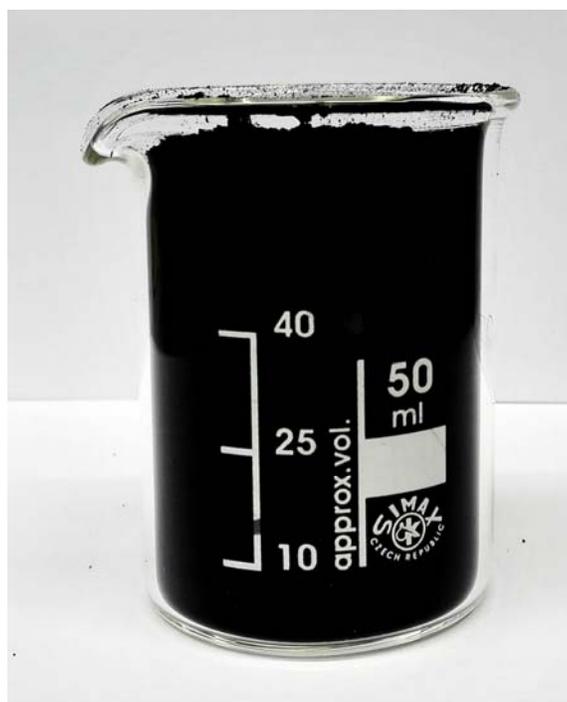

**Fig. S5.** Photograph of the product NPC-4 synthesized on a large scale of as much as ~40 g.

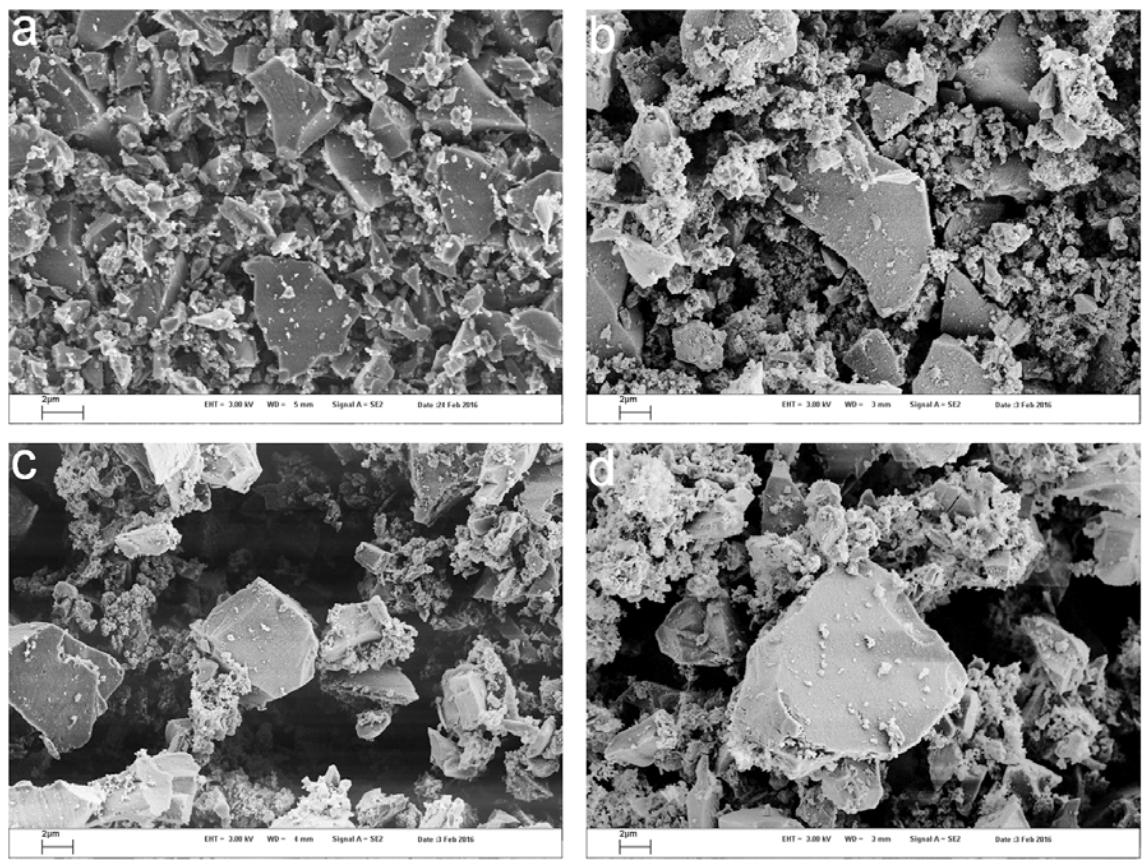

**Fig. S6**. Typical SEM images of (a) NPC-0, (b) NPC-2, (c) NPC-4, and (d) NPC-6 at a low magnification.

As we can see, NPCs comprise of many particles with a size ranging from several hundred nanometers to a couple of micrometers. In the SEM images at a low magnification, the activated samples show no obvious differences compared to the non-activated sample. This is possibly because during KOH activation the molted KOH firstly mixes well with the PIL-derived carbon and then produces micropores, which is hardly visible by SEM but easily confirmed by $N_2$ sorption measurements.

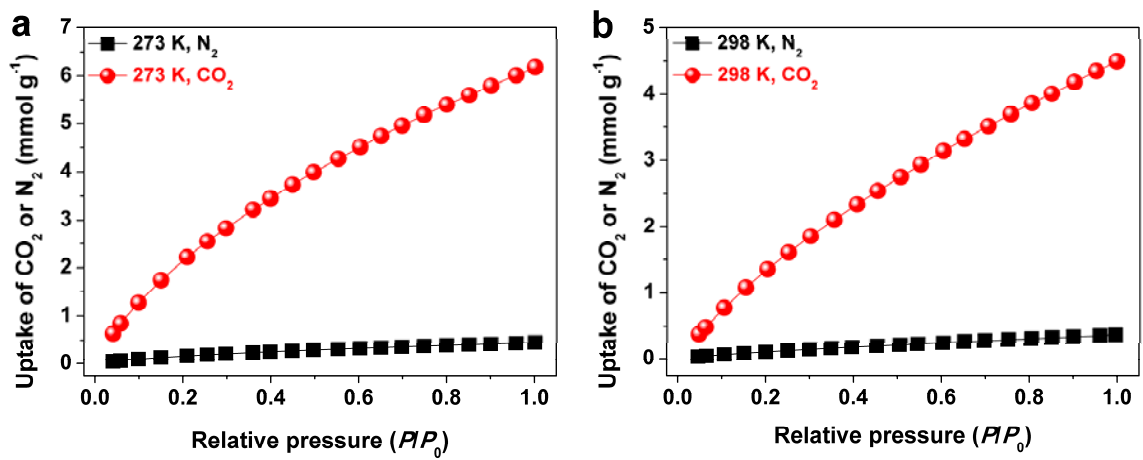

**Fig. S7.** $CO_2$ and $N_2$ adsorption isotherms of NPC-4 measured at (a) 273 K and (b) 298 K.

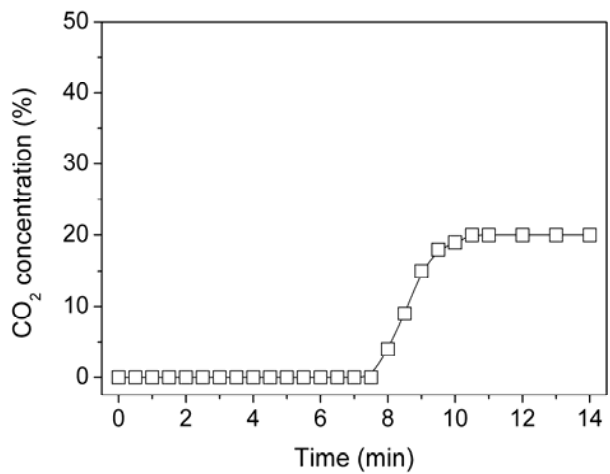

**Fig. S8.** Breakthrough curve for NPC-4 obtained at 298 K and 1 bar.

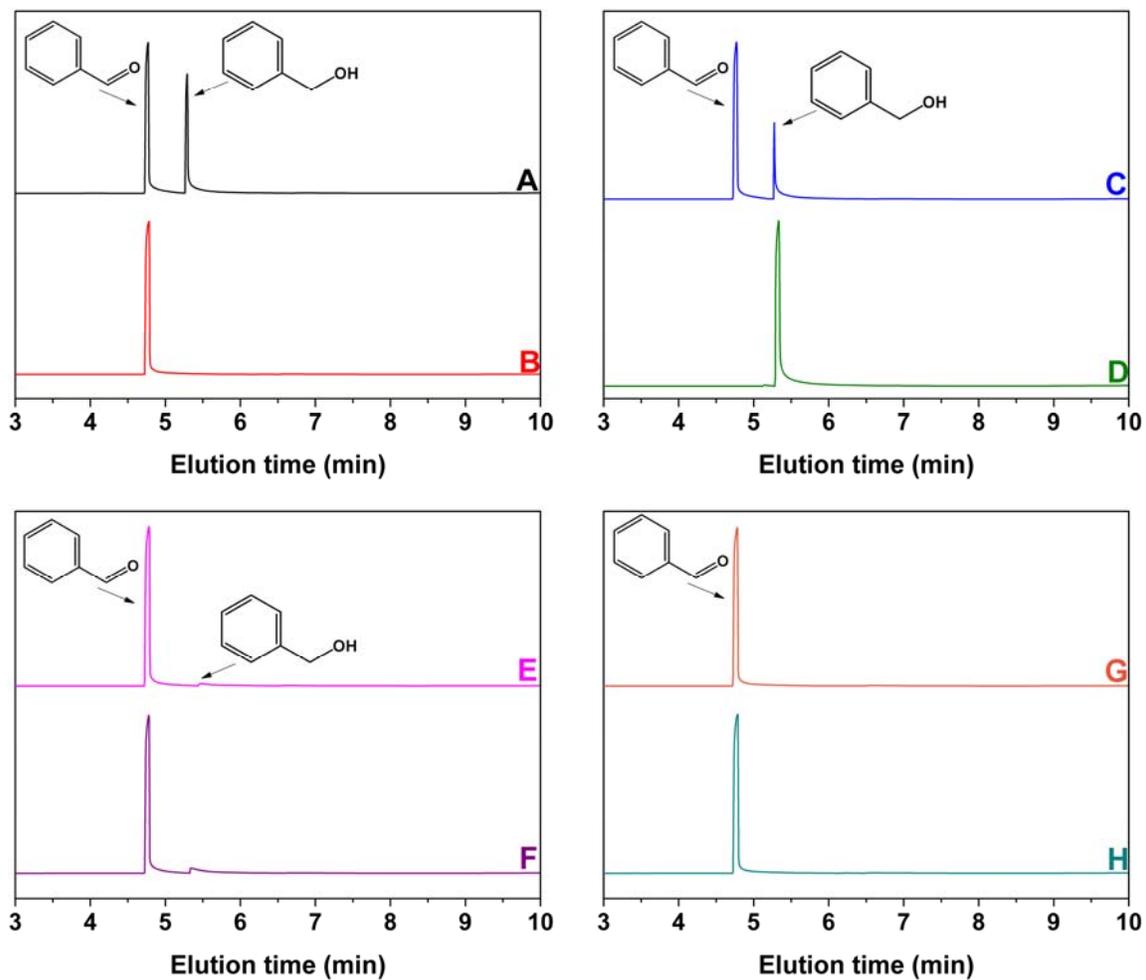

**Fig. S9.** GC-MS profiles of the product mixture from (A) entry 1, (B) entry 2, (C) entry 3, (D) entry 4, (E) entry 5, (F) entry 6, (G) entry 7, and (H) entry 8 in Table 3.

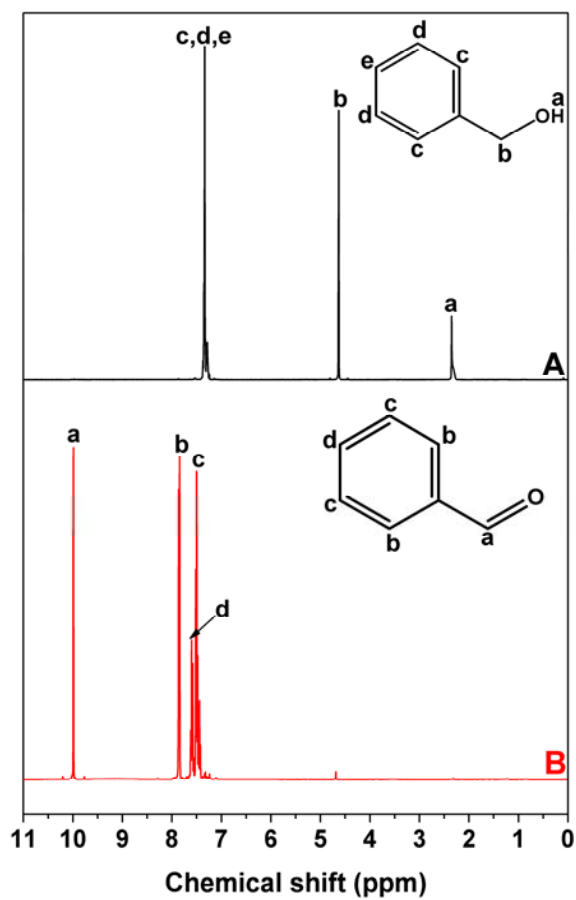

**Fig. S10.** $^1$H NMR spectra of (A) benzyl alcohol and (B) the product from entry 2 in Table 3.

## References to *SM*